\documentclass[english]{article}
\usepackage{mathpazo}
\usepackage[T1]{fontenc}
\usepackage[utf8]{inputenc}
\usepackage[a4paper]{geometry}
\usepackage{babel}
\usepackage{units}
\usepackage{graphicx}
\usepackage{setspace}
\usepackage[authoryear]{natbib}
\usepackage[unicode=true]
 {hyperref}

\makeatletter

\providecommand{\tabularnewline}{\\}

\newcommand{\lyxaddress}[1]{
\par {\raggedright #1
\vspace{1.4em}
\noindent\par}
}

\makeatother

\begin{document}

\title{Multi-scale comparative spectral analysis of satellite total solar
irradiance measurements from 2003 to 2013 reveals a planetary modulation
of solar activity and its non-linear dependence on the 11-year solar
cycle }

\author{Nicola Scafetta$^{1,2}$ and Richard C. Willson$^{1}$}

\maketitle

\lyxaddress{$^{1}$Active Cavity Radiometer Irradiance Monitor (ACRIM) Lab, Coronado,
CA 92118, USA}

\lyxaddress{$^{2}$Duke University, Durham, NC 27708 USA}
\begin{abstract}
Herein we adopt a multi-scale dynamical spectral analysis technique
to compare and study the dynamical evolution of the harmonic components
of the overlapping ACRIMSAT/ACRIM3, SOHO/VIRGO and SORCE/TIM total
solar irradiance (TSI) records during 2003.15 to 2013.16 in solar
cycles 23 and 24. The three TSI time series present highly correlated
patterns. Significant power spectral peaks are common to these records
and are observed at the following periods: $\sim0.070$ year, $\sim0.097$
year, $\sim0.20$ year, $\sim0.25$ year, $\sim0.30-0.34$ year, $\sim0.39$
year. Less certain spectral peaks occur at about $0.55$ year, $0.60-0.65$
year and $0.7-0.9$ year. Four main frequency periods at $\sim24.8$
days ($\sim0.068$ year), $\sim27.3$ days ($\sim0.075$ year), at
$\sim34$-$35$ days ($\sim0.093$-$0.096$ year) and $\sim36$-$38$
days ($\sim0.099$-$0.104$ year) characterize the solar rotation
cycle. The amplitude of these oscillations, in particular of those
with periods larger than $0.5$ year, appears to be modulated by the
$\sim$11-year solar cycle. Similar harmonics have been found in other
solar indices. The observed periodicities are found highly coherent
with the spring, orbital and synodic periods of Mercury, Venus, Earth
and Jupiter. We conclude that solar activity is likely modulated by
planetary gravitational and electromagnetic forces acting on the sun.
The strength of the sun’s response to planetary forcing depends non-linearly
on the state of internal solar dynamics: planetary-sun coupling effects
are enhanced during solar activity maxima and attenuated during minima.
\end{abstract}

\section{Introduction}

Total solar irradiance (TSI) satellite measurements are fundamental
to the investigation of solar physics and the climate change forcing
of TSI variability. TSI observations follow the solar magnetic activity
level \citep{Willson1991} and their variation therefore conforms
to the $\sim11$-year Schwabe solar cycle. The average TSI on solar
cycle time scales is sometimes referred to as the \textit{solar constant}.
TSI records are characterized by complex variability from the quasi
monthly differential solar rotation cycles to the sub-annual and annual
time scales whose origins are still unknown.

An important physical issue is whether the annual and sub-annual TSI
variability is intrinsically chaotic and unpredictable or, alternatively,
is made of a complex set of harmonics and may be predicted once a
sufficient number of constituent harmonics are identified. The latter
possibility implies solar activity forecasts and may benefit from
harmonic constituent modeling, as have the predictions of ocean tidal
levels on Earth using a set of specific solar and lunar orbital harmonics
\citep{Doodson,Kelvin}.

The harmonic constituent model hypothesis is important because it
could provide an explanation of many solar magnetic and radiative
phenomena that conventional solar physics cannot. The conventional
view of solar science is that solar magnetic and radiant variability
is intrinsically chaotic, driven by internal solar dynamics alone
and characterized by hydromagnetic solar dynamo models \citep{Tobias}.
These models cannot predict solar activity and have not been able
to explain its complex variability.

A growing body of empirical evidence suggests that solar activity
on monthly to millennial time scales may be modulated by gravitational
and magnetic planetary harmonic forces \citep[e.g.: ][]{Abreu,Brown,Charvatova,Fairbridge,Hung,Jose,Scafetta2010,Scafetta2010b,Scafetta2012a,Scafetta2012b,Scafetta2012c,Scafetta2012d,Scafetta2013,Sharp,Tan,Wilson,Wolf,Wolff}.
For example, the 11-year solar cycle appears to be bounded by the
Jupiter-Saturn spring tide oscillation period (9.93 year) and the
Jupiter orbital tide oscillation period (11.86 year) \citep{Scafetta2012c}.
The 11-year solar cycle is also in phase with major tidal resonances
generated by Venus-Earth-Jupiter system (11.07-year period) and by
Mercury-Venus system (11.08-year period) \citep{Scafetta2012d}. The
multi-decadal, secular and millennial solar oscillations appear to
be generated by beat interferences among the multiple cycles that
comprise the 11-year solar cycles \citep{Scafetta2012c}.

A recent commentary in Nature discusses the “revival” of the planetary
hypothesis of solar variation \citep{Charbonneau}. It has been pointed
out that the arguments of critics of this hypothesis \citep[e.g.: ][]{Callebaut,Smythe}
have either not been supported by empirical evidence or have based
their arguments on overly simplistic Newtonian analytical physics
\citep[e.g.: ][]{Scafetta2012c,Scafetta2012d,Scafetta2013b}.

In a previous publication \citet{ScafettaW2013b} analyzed the power
spectra of TSI records since 1992. These were compared with theoretical
power spectra deduced from the planetary orbital effects such as the
tidal potential on the sun, and the speed, jerk force and z-axis coordinate
of the sun relative to the barycenter of the solar system. The authors
found multiple evidences of spectral coherence on annual and sub-annual
scales between TSI power spectra and theoretical planetary spectra.
This suggests that TSI is modulated at specific frequencies by gravitational
and/or electromagnetic forcings linked to the revolution of the planets
around the sun.

\citet{ScafettaW2013b} found a TSI signature of the 1.092-year Earth-Jupiter
conjunction cycle. The TSI oscillation was found to be particularly
evident during the maximum of solar cycle 23 (1998-2004) and in phase
synchronization with the Earth-Jupiter conjunction cycle that predicts
an enhanced effect when the Earth crosses the Sun-Jupiter conjunction
line. The cause was postulated to be a slightly brighter side of the
Sun facing Jupiter because that side would be the focus of enhanced
planetary-solar couplings, both gravitational and electromagnetic.
These forces exerted by Jupiter on the Sun are the strongest of all
the planets. When the Earth crosses the Sun-Jupiter conjunction line
it adds to Jupiter’s planetary-solar coupling effects and sensors
on Earth satellites should receive a stronger TSI signal. This planetary-solar
coupling effect generate the \textasciitilde{}1.092-year cycle in
the TSI record.

The 1.092-year cycle signature detected by the satellite TSI observations
is enhanced during solar activity maxima and attenuated during solar
minima \citep{ScafettaW2013b} suggesting complex, non-linear responses
of solar internal dynamics to planetary forcings. Here we study the
dynamical evolution of the harmonic characteristics of the TSI observations
on annual and sub annual time scales. A multiscale dynamical spectral
analysis technique is proposed and used to reveal non-stationary changes
in dynamical patterns in a sequence. The technique is used to determine
whether major background harmonics exist that correspond to basic
planetary harmonics such as the spring, orbital and synodic periods
among the planets.

\section{Total Solar Irradiance Data}

The daily average TSI measurements were collected during the last
decade by three independent science teams: ACRIMSAT/ACRIM3 \citep{Willson2003},
SOHO/VIRGO \citep{Frohlich} and SORCE/TIM \citep{Kopp2005a,Kopp2005b}.
Cross-comparison of the three independent TSI records reduces interpretation
errors due to measurement uncertainties. Dynamical patterns common
to the three TSI records are sought to increase the certainty of their
physical origins.

ACRIM3 results have been adjusted using algorithm updates and corrections
for scattering and diffraction found in recent testing at the LASP/TRF
--TSI Radiation Facility (TRF) of the Laboratory for Atmospherics
and Space Physics (LASP)-- \citep{Willson28}. Similar LASP/TRF corrections
were recently found for the VIRGO results and they are now reported
on an updated scale \citep{Frohlich201329}. The updated ACRIM3, VIRGO
and TIM results agree closely in scale and variability with an average
value during the 2008-2009 solar activity minimum near $1361$ $W/m^{2}$.

\begin{figure}[!t]
\includegraphics[width=1\textwidth]{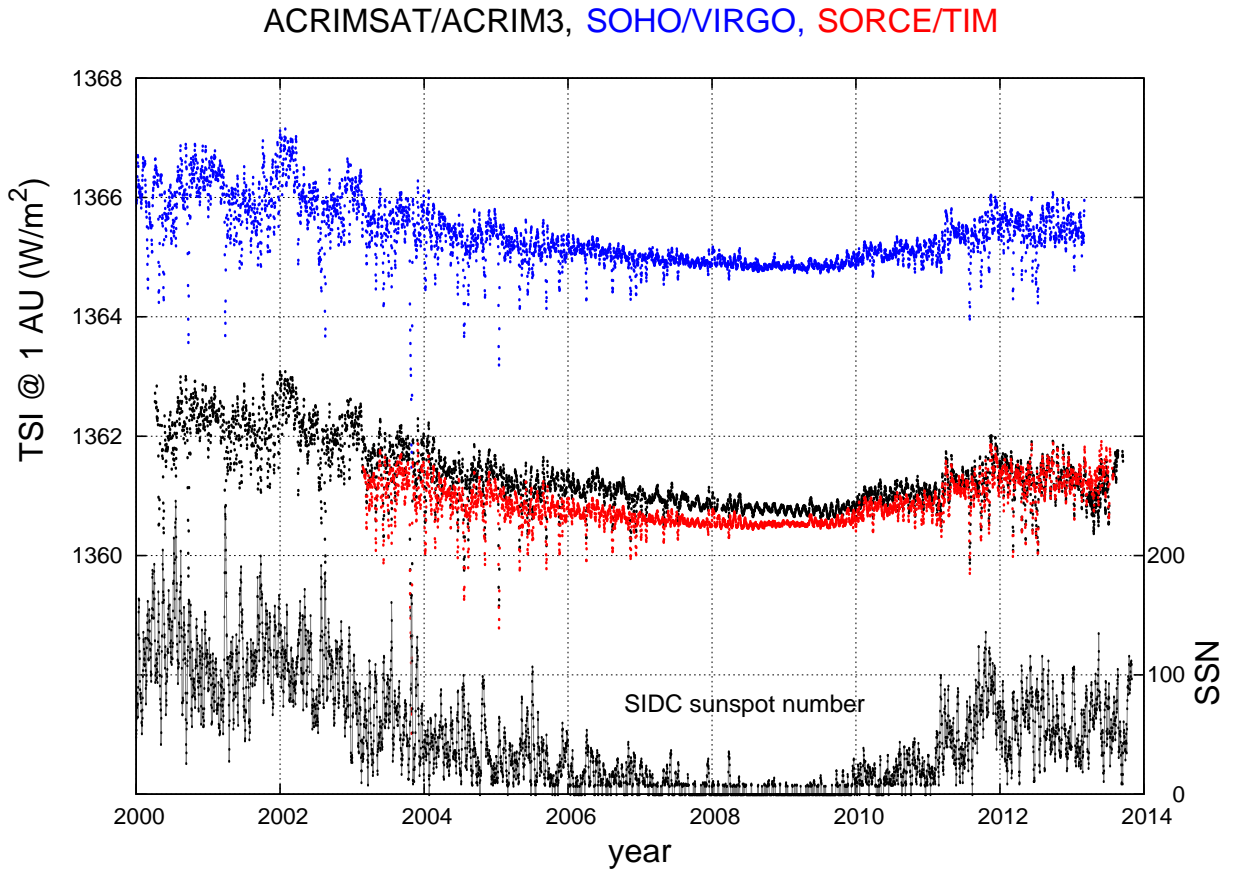}\caption{Comparison of ACRIMSAT/ACRIM3 (black), SOHO/VIRGO (blue) and SORCE/TIM
(red) total solar irradiance records versus the SIDC daily sunspot
number (gray) record. ACRIM3 is recalibrated with updated sensor degradation
and algorithm LAPS/TRF corrections for scattering, diffraction and
SI scale. VIRGO does not include yet the LASP/TRF scaling corrections. }
\end{figure}

The ACRIMSAT/ACRIM3, SOHO/VIRGO and SORCE/TIM TSI records since 2003
are shown in Figure 1. For comparison, Figure 1 also depicts the daily
sunspot number record from the Solar Influences Data Analysis Center
(SIDC).

Note that Figure 1 shows the most recent SOHO/VIRGO record available
that does not yet include the LASP/TRF scaling corrections. Thus,
it is more significant to compare the three TSI records as percentage
variation about successive two year periods as depicted in Figures
2 and 3.

\begin{figure}[!t]
\includegraphics[width=1\textwidth]{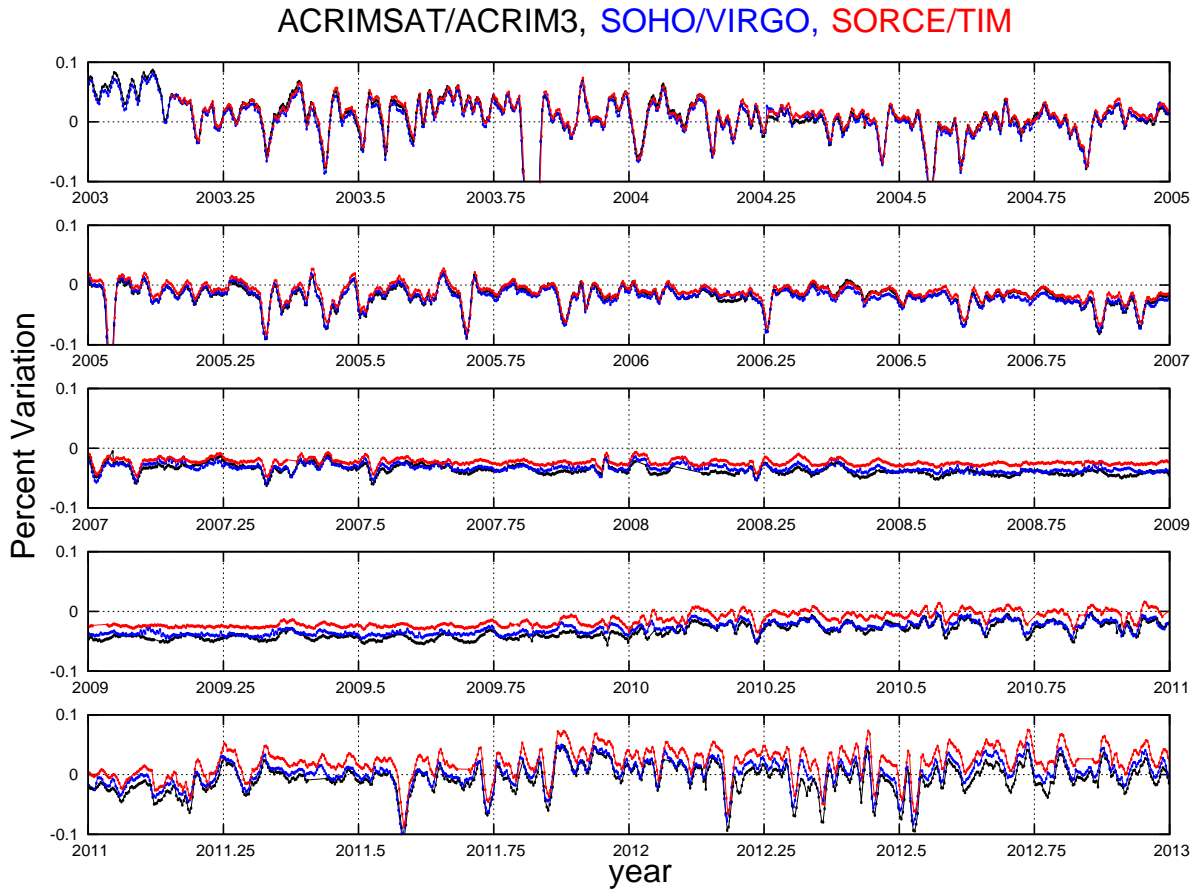}\caption{Percent variation comparison of ACRIMSAT/ACRIM3 (black), SOHO/VIRGO
(blue) and SORCE/TIM (red) total solar irradiance records. The scale
of the two year segments is constant to highlight the divergent trend
of the TIM results relative to those of the ACRIM3 and VIRGO experiments.
The records are cross-scaled during the initial two-weeks period 2003.15-2003.19.}
\end{figure}

Figure 2 uses a constant scale for each two-year period to demonstrate
the progressive divergence of TIM relative to ACRIM3 and VIRGO results.
The three records are scaled during the initial common two-week period
(2003.15-2003.19). The close agreement of all three satellite experiments'
results in 2003 was followed by continuous divergence of TIM results
from those of ACRIM3 and VIRGO through 2013 when the difference reached
\textasciitilde{} 200 ppm.

The most likely cause of the divergence, based on previous satellite
TSI observational experience, is in-fl{}ight sensor degradation calibration
error. The close agreement of ACRIM3 and VIRGO results, which is more
evident in Figures 2-3, indicates that (1) an over-correction of TIM
sensor degradation is the most likely explanation. However, the cause
could also be (2) a combination of degradation uncertainty by all
three or (3) may be within the uncertainty of the self-degradation
calibration capabilities of these instruments. The long term traceability
of TSI satellite results, achieved through in-fl{}ight self-calibration
of degradation, is an important area of continuing research for the
climate TSI database.

\begin{figure}[!t]
\includegraphics[width=1\textwidth]{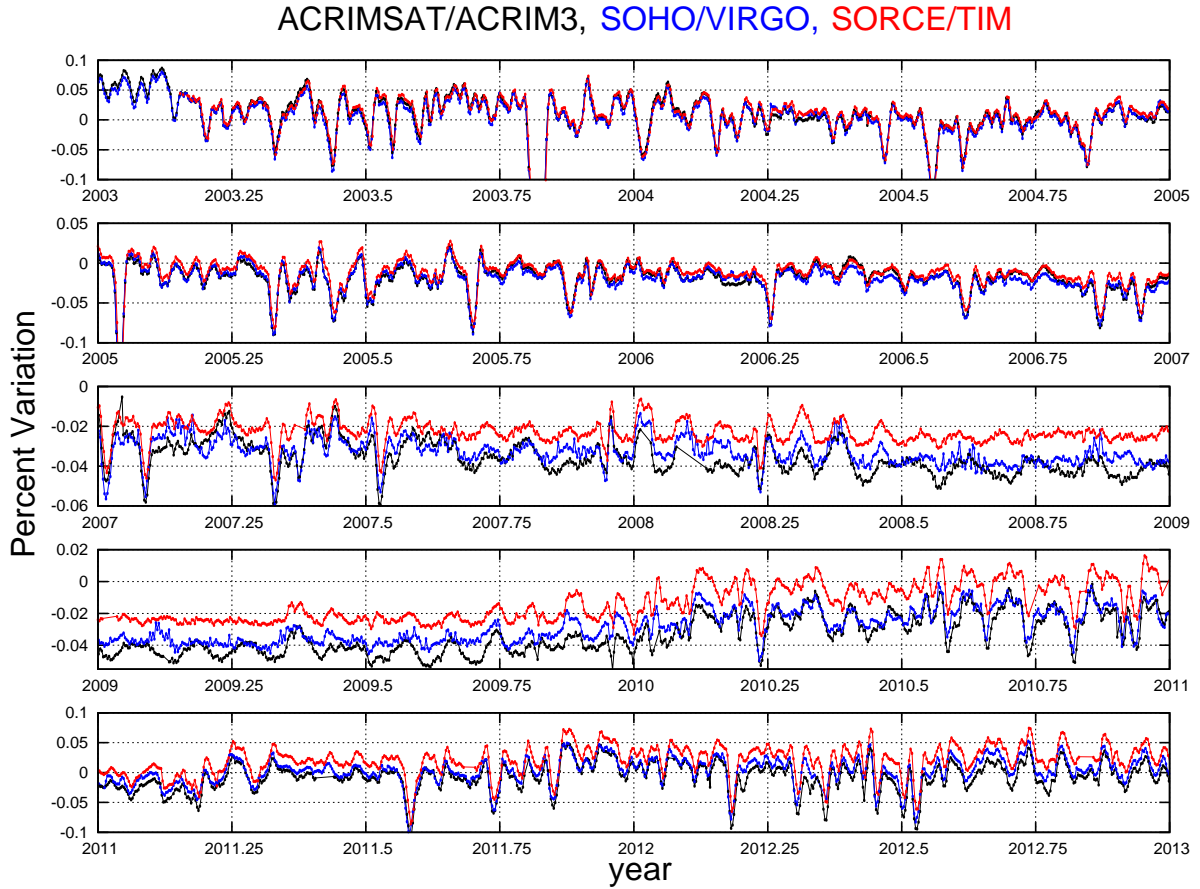}\caption{Percent variation comparison of ACRIMSAT/ACRIM3 (black), SOHO/VIRGO
(blue) and SORCE/TIM (red) total solar irradiance records. The scale
of the two year segments is varied to highlight the detailed similarity
of the variability of all three TSI records. The records are cross-scaled
during the initial two-weeks period 2003.15-2003.19.}
\end{figure}

Figure 3 uses a variable scale on each two-year segment to provide
maximum visibility of the TSI variations for each sensor. It can be
clearly seen that ACRIM3, VIRGO and TIM detect nearly all the same
variations. TIM appears to see them with slightly lower amplitudes.
During the quietest magnetic activity part of the solar minimum period
(2008.7–2009.3) there is a near absence of variations in the TIM record
while VIRGO sees some of the variability detected by ACRIM3 during
this time but at lower amplitudes. Lower sensitivities of VIRGO and
TIM sensors is likely responsible for these differences.

\section{TSI power spectrum comparison}

Power spectrum evaluations of the TSI records are shown as Figures
4, 5 and 6. In the following two subsections we analyze the multi-montly
scale (0.1-1.1 year) and the solar differential rotation scale (22-40
days).

\begin{figure}[!t]
\includegraphics[width=1\textwidth]{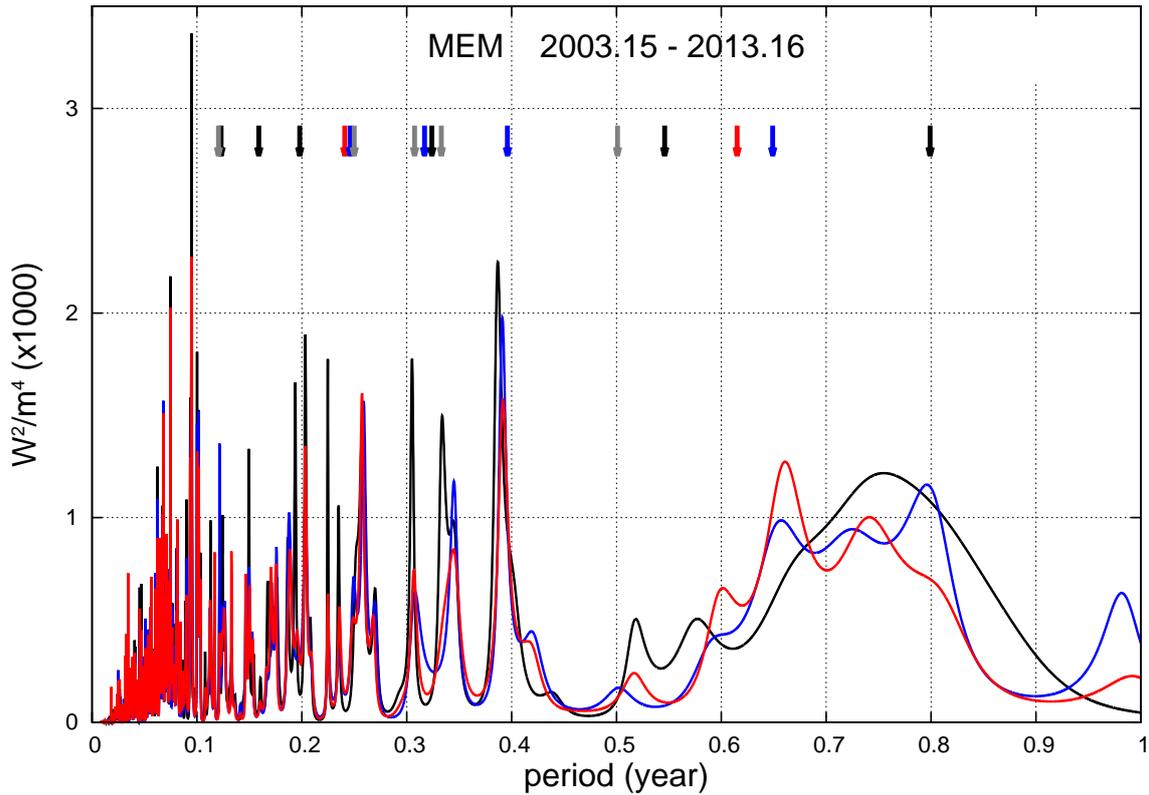}\caption{Maximum entropy method (MEM) power spectrum comparison of ACRIMSAT/ACRIM3
(black), SOHO/VIRGO (blue) and SORCE/TIM (red) total solar irradiance
records using the data from 2003.15 to 2013.16. The colored arrows
at the top of the figure indicate the major theoretically expected
planetary frequencies from Mercury, Venus, Earth and Jupiter, which
are reported in Table 1. The red color indicates the orbital periods,
the black color indicates the spring periods, the blue color indicates
the synodic periods and the gray color indicates the harmonics of
the orbital periods listed in Table 1. }
\end{figure}

\subsection{Analysis of the 0.1-1.1 year period range }

Figure 4 shows the maximum entropy method (MEM) power spectrum evaluation
\citep{Press} of the ACRIM3, VIRGO and TIM TSI records during the
ten-year period from 2003.15 to 2013.16. The power spectra are plotted
as a function of the period ($T=1/f$) up to 1 year. The fi{}gure
shows that the three records produce nearly all the same spectral
peaks indicating that the observed variations in TSI are definitively
solar in origin. The spectral amplitude of the peaks for ACRIM3 record
is nearly always higher than that observed for VIRGO and TIM, indicating
a higher sensitivity of ACRIM3 instrumentation for TSI variability.
This sensitivity difference is supported as well by the fact that
the TIM and VIRGO records present slightly smoothed and attenuated
patterns relative to those of ACRIM3. The major spectral peaks are
highlighted in the figure, and occur at the following approximate
periods: $\sim0.070$ year, $\sim0.095$ year, $0.20$ year, $0.25$
year, $0.30-0.34$ year, $0.39$ year and $0.75-0.85$ year; more
uncertain peaks occur at about $0.60-0.65$ year.

The above periods are found among the major planetary harmonics related
to the orbital, synodic and spring periods for the planets. Tables
1 reports these periods with their uncertainty and range during 2003-2013
for the four major tidal-causing planets (Mercury, Venus, Earth and
Jupiter) \citep{Scafetta2012d}. Table 2 shows other theoretically
expected periods related to the other planets as well. The major orbital,
synodic and spring periods listed in Table 1 are indicated with colored
arrows at the top of Figure 4: red indicates orbital periods, black
indicates spring periods, blue indicates the synodic periods and gray
indicates the harmonics of the orbital periods listed in Table 1.
The additional planetary frequencies listed in Table 2 likely have
only minor TSI effects and are not explicitly called out in Figure
4; we report these additional frequencies for completeness. Although
there is currently a defi{}cit of specifi{}c physical mechanisms to
explain planet-Sun interactions, these minor frequencies may also
be found in solar records.

\citet{ScafettaW2013b} found similar frequencies using theoretical
equations deduced from the ephemerides of the planets such as the
tidal potential on the Sun and the speed, jerk force and z-axis coordinate
of the Sun relative to the barycenter of the solar system. Statistical
tests based on Montecarlo simulations using red-noise generators for
TSI synthetic records established that the probability of randomly
finding a dynamical sequence manifesting a spectral coherence with
the (orbital, spring and synodic) planetary theoretical harmonics
equal or larger than that found among the TSI satellite frequencies
and the planetary harmonics is less than 0.05\% \citep{ScafettaW2013b}.

A comparison between the spectral peaks shown in Figures 4 and the
colored arrows indicating the major expected planetary frequency peaks
shows a clear coherence among the TSI and the planetary harmonics
on multiple scales, in particular for the periods from 0.1 year to
0.5 year and for the 0.8-year periodicity. The three planetary periods
at about 0.55 year and between 0.6 year and 0.65 year do not appear
equally evident in the TSI results.

\begin{figure}[!t]
\includegraphics[width=1\textwidth]{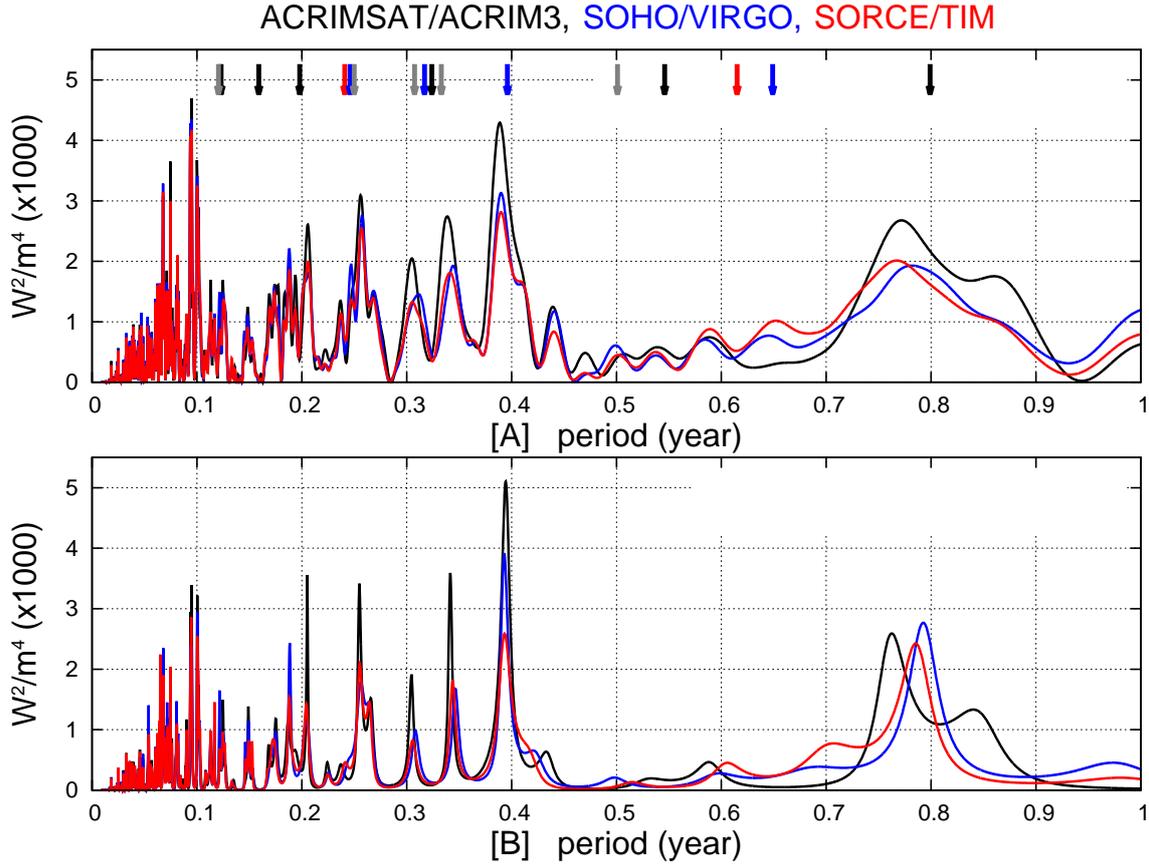}\caption{Power spectrum comparison of ACRIMSAT/ACRIM3 (black), SOHO/VIRGO (blue)
and SORCE/TIM (red) total solar irradiance records using the data
from 2003.15 to 2011.00. {[}A{]} The periodogram is used; {[}B{]}
the maximum entropy method (MEM) is used. The colored arrows at the
top of the figure indicate the major theoretically expected planetary
frequencies from Mercury, Venus, Earth and Jupiter, which are reported
in Table 1. The red color indicates the orbital periods, the black
color indicates the spring periods, the blue color indicates the synodic
periods and the gray color indicates the harmonics of the orbital
periods listed in Table 1. }
\end{figure}

As discussed in the Introduction, the response of the Sun to external
planetary forcing may be non-linear with some frequencies enhanced
by internal solar dynamics during specifi{}c periods (e.g. solar maxima)
and attenuated during others (e.g. solar minima). Indeed, changing
the analyzed period as done in Figure 5 (we used the data from 2003.15
to 2011) produces some differences relative to the results depicted
in Figure 4. For example the amplitudes of the peaks are different
although their frequency position is fairly unchanged. This demonstrates
that non-linear mechanisms are regulating the phenomenon. Section
4 addresses the non-linear dynamical evolution of the TSI patterns
more systematically.

\begin{figure}[!t]
\includegraphics[width=1\textwidth]{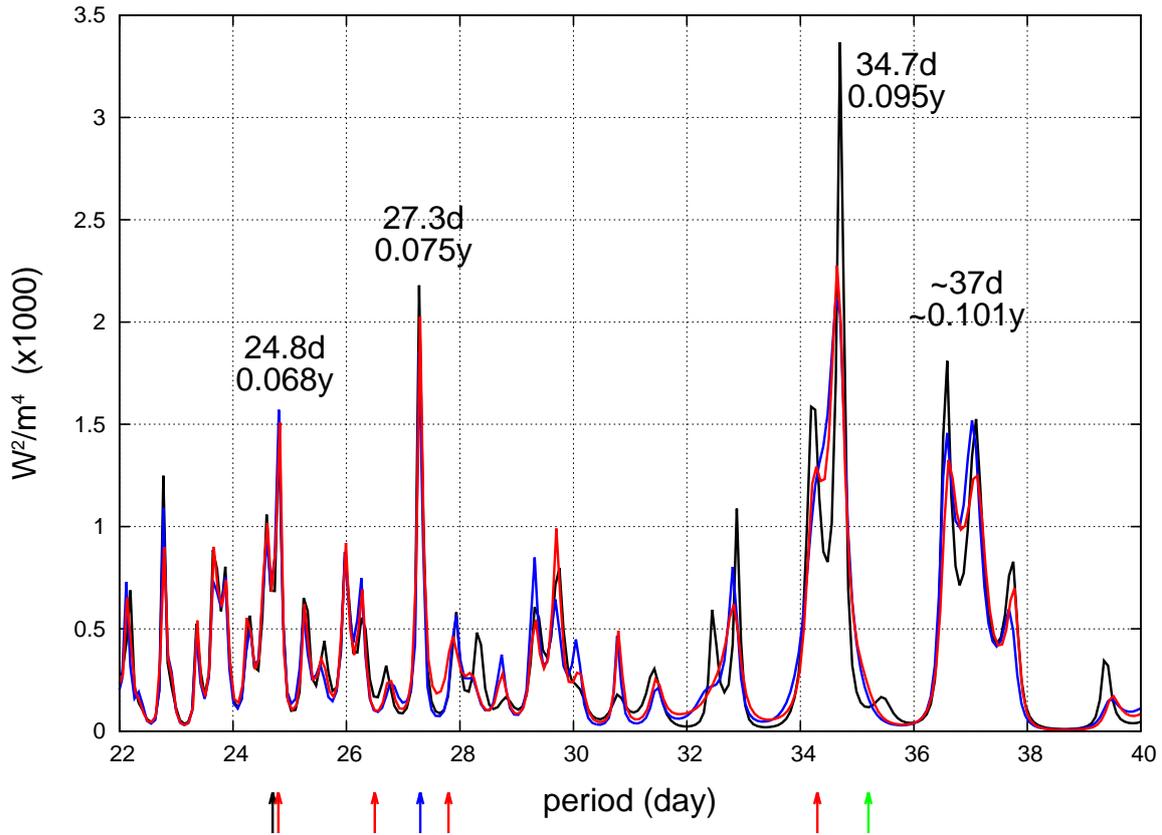}\caption{Magnification of the period range from 22 to 40 days depicted in Figure
4 which is associated with the solar differential rotation scale.
The arrows at the bottom depicts the solar rotation cycles reported
in Table 3. }
\end{figure}

\subsection{Analysis of the 22-40 day period range associated to the solar differential
rotation}

Figure 6 magnifies Figure 4 between 22 and 40 days. This range corresponds
to the differential solar rotation period band. Figure 6 clearly shows
a spectral peak at \textasciitilde{}27.3 days (0.075 year) \citep{Willson1999}.
This corresponds to the synodic period between the well-known Carrington
solar rotation (\textasciitilde{}25.38 days) and the Earth's orbital
period (\textasciitilde{}365.25 days). The Carrington period roughly
corresponds to the solar rotation period at a latitude of $26^{o}$
from the sun's equatorial plane, which is the average latitude of
sunspots and corresponding magnetic solar activity \citep{Bartels},
as seen from the Earth.

Figure 6 also reveals spectral peaks at $\sim24.8$ days ($\sim0.068$
year), at $\sim34$-$35$ days ($\sim0.093$-$0.096$ year) and $\sim36$-$38$
days ($\sim0.099$-$0.104$ year) suggesting that the sidereal equatorial
and polar solar rotation cycles would be also detected in TSI records.
However, the presence of these cycles in the TSI records could imply
that the solar orientation relative to space also modulates solar
activity. A possible explanation of these spectral peaks could require
a planetary influence on the sun.

Assuming that the Jupiter's side of the sun is the focus of higher
solar activity \citep{ScafettaW2013b} it is possible to interpret
the $\sim24.8$ days ($\sim0.0679$ year) cycle as the synodic period
between Jupiter's sidereal orbital period (\textasciitilde{}4332.6
days = \textasciitilde{}11.862 years) and the solar equatorial rotation
period. The latter is estimated to be $\sim24.7$ days ($\sim0.0675$
year) during the period analyzed here from 2003 to 2013. Using this
estimate, additional planetary synodic cycles with the solar rotation
are calculated at: $\sim26.5$ days ($\sim0.0725$ year), the synodic
solar equatorial rotation period relative to Earth; $\sim27.75$ days
($\sim0.0760$ year), the synodic solar equatorial rotation period
relative to Venus; and $\sim34.3$ days ($\sim0.0940$ year), the
synodic solar equatorial rotation period relative to Mercury. See
Table 3.

The $\sim34.3$-day Mercury-sun synodic period fits the TSI spectral
peak at $\sim34$-$35$ days, a period that corresponds also to the
high latitude solar differential rotation rate. However, the theoretical
synodic spectral peaks at $\sim26.5$ days and $\sim27.75$ days do
not appear in Fig. 5. This would suggest that solar asymmetry causes
a TSI enhancement as the sun's more sensitive region orientates only
toward Jupiter and Mercury.

\begin{figure}[!t]
\includegraphics[width=1\textwidth]{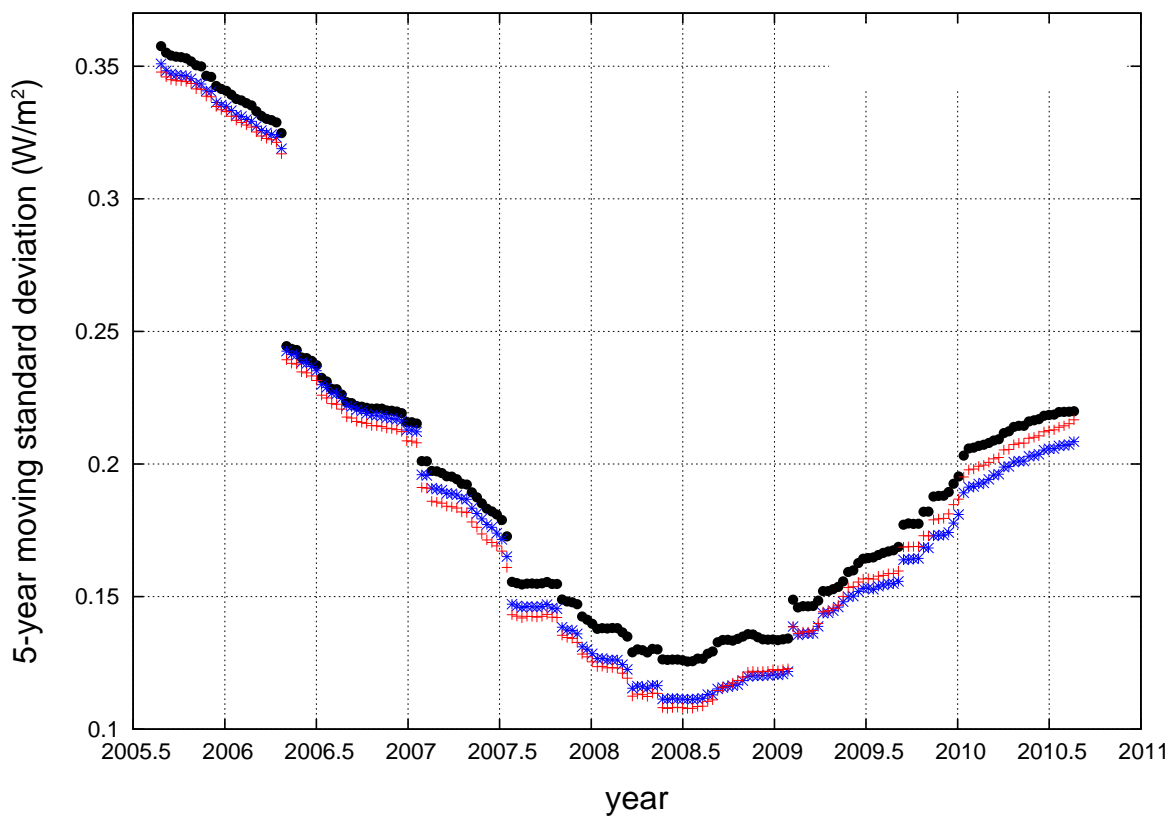}\caption{5-year moving standard deviation function $\sigma_{5}(t)$ of the
high-pass filtered ACRIMSAT/ACRIM3 (black), SOHO/VIRGO (blue) and
SORCE/TIM (red) total solar irradiance records depicted in Figure
6.}
\end{figure}

Mercury may have a strong effect because Mercury is the closest planet
to the sun. After Jupiter, Mercury induces on the Sun the second largest
tidal amplitude cycle related to a planetary orbit due to its large
eccentricity ($e=0.206$) and low inclination to Sun's equator ($3.38^{o}$)
\citep[Figure 8]{Scafetta2012d}. Moreover, the theoretical $\sim34.3$
day Mercury-sun synodic period is near a 2/5 resonance with Mercury's
orbital period ($\sim88$ days) or $\sim35.2$ days ($\sim0.096$
year). This close resonance may favor dynamical synchronization and
amplification in solar dynamics and explain the wide, strong TSI spectral
peak around $\sim34$-$35$ days that appears bounded by Mercury's
two theoretical frequencies as Figure 6 shows.

Thus empirical evidence suggests that the differential solar rotation
may be synchronized to the synodic cycles between the solar equatorial
rotation and the two theoretically most relevant tidal planets: Jupiter
and Mercury. Further investigation of the solar rotation period band
requires a dedicated investigation that is left to another work.

\begin{figure}[!t]
\includegraphics[width=1\textwidth]{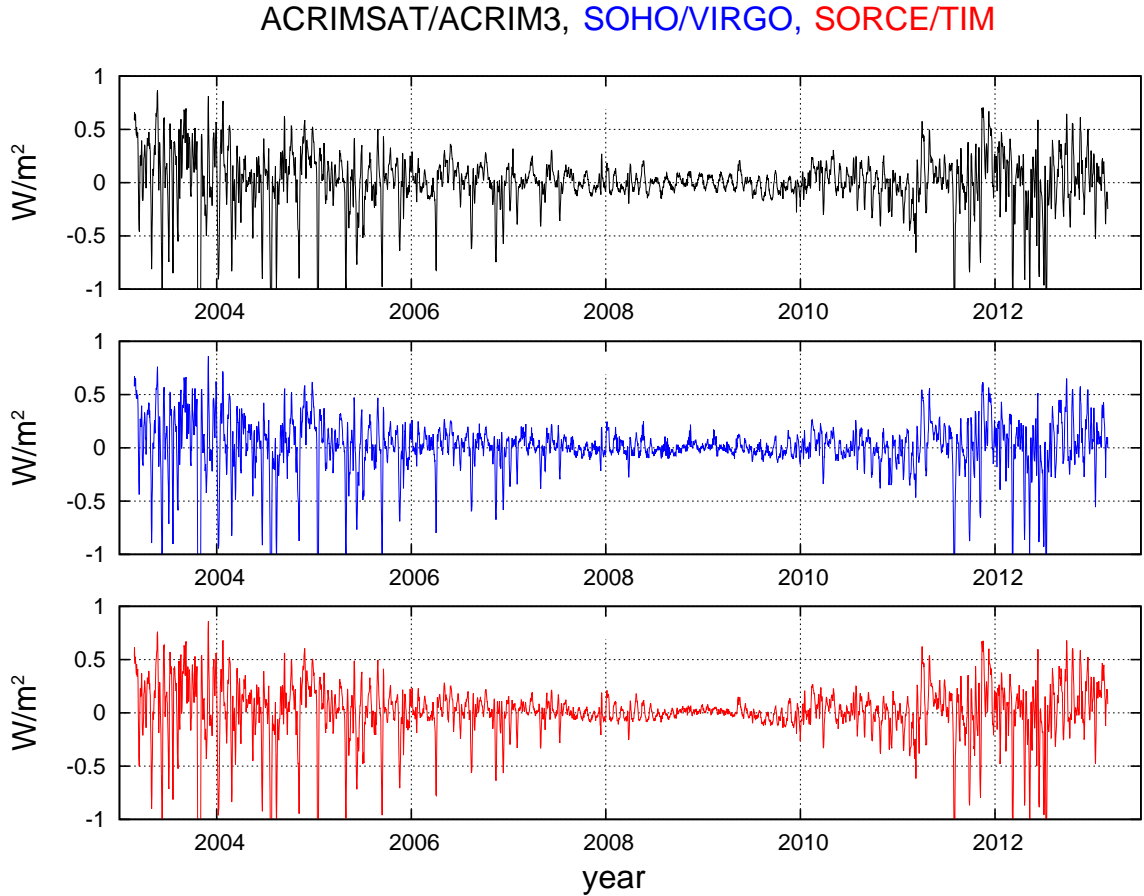}\caption{FFT 2-year high-pass filtered component of the ACRIMSAT/ACRIM3 (black),
SOHO/VIRGO (blue) and SORCE/TIM (red) total solar irradiance records.}
\end{figure}

\section{Multi-scale comparative spectral analysis}

Multi-scale dynamical spectral analysis diagrams for the three TSI
records were constructed as follows. We consecutively calculated normalized
power spectrum functions using MEM, which produces sharp peaks and
it is less affected by leakage artifacts. We processed the TSI records
after high-pass filtering to eliminate time scale variations longer
than 2 years. Figure 7 depicts the Fast Fourier Transform (FFT) 2-year
high-pass filtered components of the three TSI records. We analyzed
consecutive 5-year moving centered windows of the data (for example,
the results centered in 2006 refer to the 5-year period from 2003.5
to 2008.5).

Figure 8 shows the 5-year moving standard deviation functions, $\sigma_{5}(t)$,
of the high-pass filtered TSI records that were used for local normalization
of the MEM functions. During the solar minimum $\sigma_{5}(t)$ is
attenuated relative to the solar cycle 23 and 24 maxima in all three
TSI records.

\begin{figure}[!t]
\includegraphics[width=1\textwidth,height=0.8\textheight]{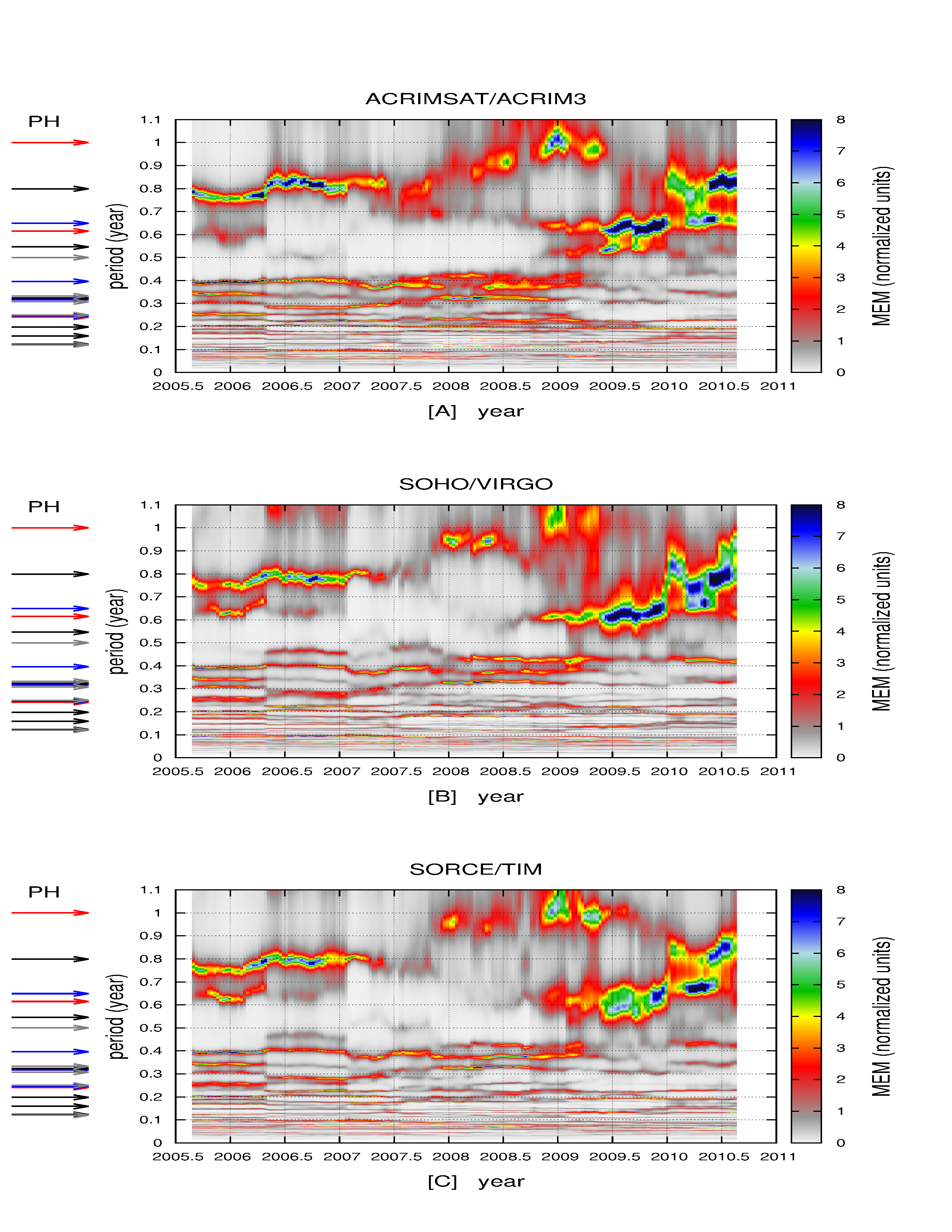}\caption{Moving window power spectrum comparison of {[}A{]} ACRIMSAT/ACRIM3;
{[}B{]} SOHO/VIRGO; {[}C{]} SORCE/TIM total solar irradiance records.
The maximum entropy method (MEM) is used. The colors represent the
spectral strength in variance normalized units (x100), with the blue-black
regions representing the strongest spectral peaks. The colored arrows
at the left of the diagrams indicate the theoretically expected frequencies
of the most significant planetary harmonics (PH) obtained from Mercury,
Venus, Earth and Jupiter, which are reported in Table 1. The red color
indicates the orbital periods, the black color indicates the spring
periods, the blue color indicates the synodic periods and the gray
color indicates the harmonics of the orbital periods listed in Table
1. }
\end{figure}

The multi-scale comparative spectral analysis diagrams are depicted
in Figure 9 within the period range 0 to 1.1 year. Figure 10 magnifies
the period range from 0.10 year to 0.45 year. The diagrams were obtained
calculating MEM curves for a 5-year moving centered window and plotting
it in a column using colors to represent the strength of the spectral
function. For example, the colored column above the year 2006 corresponds
to the MEM power spectrum of the data covering the 5-year period from
2003.5 to 2008.5. The presence of harmonics even when attenuated during
solar minimum is emphasized by the colored column of Figures 9 and
10, which shows a spectrum normalized by the variance $\sigma_{5}(t)$
of the data during the analyzed 5-year interval.

Figure 9 shows that even after normalization the amplitude of some
frequencies depends strongly on the strength of solar cycle activity.
TSI oscillation variability is seen to be larger during solar maxima
and smaller during solar minima. Major peaks (blue-back color) are
observed for the same periodicities seen in Figures 4 and 5, indicated
by arrows on the left. The spectral peaks are relatively stable as
the 5-year window moves in time. The stationarity of these spectral
lines increases for periods below 0.5 year. The peaks near 0.6-0.7
year and 0.8 year are attenuated or disappear during solar cycle 23-24
minimum (\textasciitilde{} 2006.75 to 2008.75). The strong periodicities
near 0.8 year are attenuated or disappear during 2008-2009.25. In
particular the peak at 0.6-0.65 year is clearly visible before 2006.5
and after 2008.75 in all three diagrams.

Some differences are also seen in the three panels of Figures 9 and
10. The ACRIM3 panel is the most colorful, indicating the highest
detection of variability, and TIM is the least (corresponding to the
standard deviation variability depicted in Figure 8). Because the
calculations are the same for all three TSI records this implies that
the spectral peaks detected by ACRIM3 are generally stronger that
those detected by the other two experiments, providing another confirmation
that ACRIM3 sensors are more sensitive than VIRGO and TIM, recording
stronger signals on multiple scales.

\begin{figure}[!t]
\includegraphics[width=1\textwidth,height=0.8\textheight]{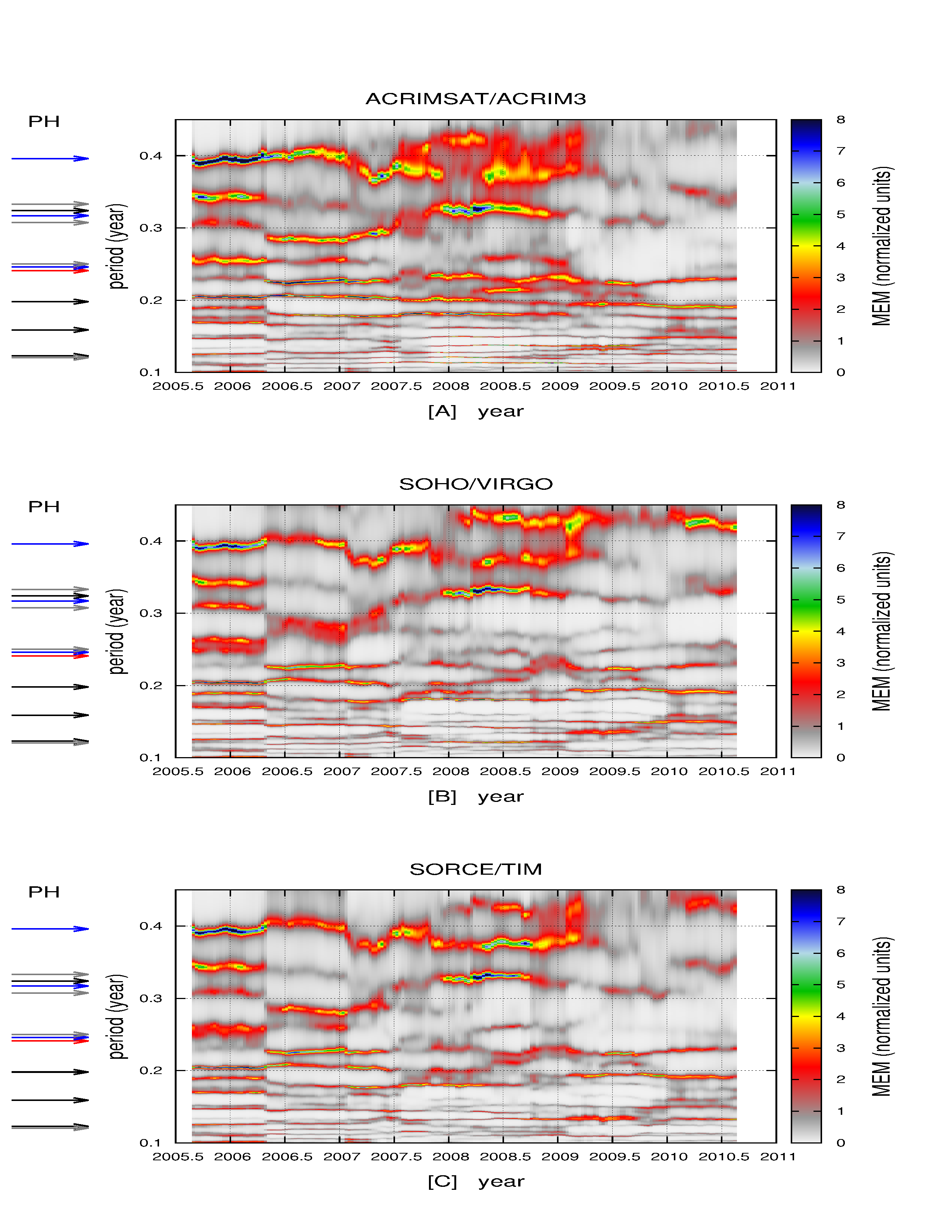}\caption{Magnification of Fig. 9 within the frequency period range from 0.10
year to 0.45 year. }
\end{figure}

\section{Discussion and Conclusions}

ACRIMSAT/ACRIM3, SOHO/VIRGO and SORCE/TIM TSI records overlap since
2003.15 and are found to be closely correlated with each other. Including
the LASP/TRF calibration corrections for both ACRIM3 and VIRGO, all
three records present a similar TSI average at about $1361$ $W/m^{2}$.
Figure 1 still depicts the SOHO/VIRGO record at the uncorrected scale
(at about $1365$ $W/m^{2}$ ) since the VIRGO updated record is not
currently available.

Power spectrum and multi-scale dynamical spectral analysis techniques
have been used to study the physical properties of these data. We
found that TSI is modulated by major harmonics at: $\sim0.070$ year,
$\sim0.097$ year, $\sim0.20$ year, $\sim0.25$ year, $\sim0.30-0.34$
year, $\sim0.39$ year; the peaks occurring at $\sim0.55$ year, $\sim0.60-0.65$
year and $\sim0.7-0.9$ year appear to be amplified during solar activity
cycle maxima and attenuated during the minima.

Other researchers have studied the fast oscillations of alternative
solar indices and found results compatible with ours. \citet{Rieger}
found that an index of energetic solar flare events presents a major
variable oscillation with a period of about 154 days (0.42 year).
Similarly \citet{Verma} found a 152-158 day (0.41-0.43 year) periodicity
in records of solar nuclear gamma ray flares and sunspots. This period
approximately corresponds to the Mercury-Venus synodic cycle ($\sim0.4$
year), which is quite evident in Figures 4 and 5, and may slightly
vary in time as shown in Figures 9 and 10. \citet{Pap} analyzed a
number of solar indices (ACRIM-1 TSI, 10.7 cm radio flux, Ca-K plage
index, sunspot blocking function and UV flux at $L\alpha$ and MgII
core-to-wing ratio) and found major spectral peaks at about 51 days
($\sim0.14$ year), 113-117 days (0.30-0.32 year), 150-157 days (0.41-0.43
year), 227 days ($\sim0.62$ year) and 240-330 days (0.65-0.90 year).
\citet{Caballero} analyzed the fluctuations detected in high-altitude
neutron monitor, solar and interplanetary parameters. \citet{Kilcik}
analyzed periodicities in solar flare index for solar cycles 21-23.
\citet{Tan} analyzed the solar microwave emission flux at frequency
of 2.80 GHz (F10.7) and the daily relative sunspot number (RSN) from
1965-01-01 to 2011-12-31. These three studies revealed major periodicities
within these period ranges: 53-54 days (0.14-0.15 year); 85-90 days
(0.23-0.25 year); 115-120 days (0.31-0.33 year); 140-150 days (0.38-0.41
year); 230-240 days (0.62-0.66 year); 360-370 days (0.98-1.02 year);
395-400 days (1.08-1.10 year). The periodicity ranges found above
correspond well to those found in the TSI satellite records as shown
in Figures 4, 5, 9 and 10, and correspond to major (orbital, spring
and synodic) planetary harmonics as reported in Tables 1 and 2.

Four main high frequency periods at $\sim24.8$ days ($\sim0.068$
year), $\sim27.3$ days ($\sim0.075$ year), at $\sim34$-$35$ days
($\sim0.093$-$0.096$ year) and $\sim36$-$38$ days ($\sim0.099$-$0.104$
year) characterize the differential solar rotation. The $\sim27.3$
days ($\sim0.075$ year) period is the well known Earth's synodic
period with the Carrington solar rotation period ($\sim25.38$ days).
The interpretation of the other cycles is uncertain. Perhaps the $\sim24.8$
days ($\sim0.068$ year) and \textasciitilde{}34-35 days ($\sim0.093$-$0.096$
year) cycles are the synodic cycles between the equatorial solar rotation
cycle and the orbit of Jupiter and Mercury respectively. The latter
could also be synchronized to the 2/5 resonance of the Mercury orbital
period of $\sim35.2$ days ($\sim0.096$ year). The $\sim36$-$38$
days ($\sim0.099$-$0.104$ year) may be the upper bound of the polar
differential solar rotation as seen from the Earth.

In conclusion solar activity appears to be characterized by specific
major theoretical harmonics, which would be expected if the planets
are modulating it. Mercury, Venus, Earth and Jupiter would provide
the most modulation within the studied time scales \citep{Scafetta2012c,Scafetta2012d,Tan}.
If these planets are modulating solar activity via gravitational and/or
electromagnetic forces, although the physical mechanisms are still
unknown, the harmonics referring to the spring, orbital and synodic
periods among the planets should be present in the TSI records as
well. The planetary harmonics reported in Tables 1, and 3, computed
using the orbital periods of four theoretically most relevant planets
(Mercury, Venus, Earth and Jupiter) correspond very closely to the
harmonics observed in the TSI records (see Figures 4, 5, 6, 9 and
10).

Our fi{}ndings support the hypothesis that planetary forces are modulating
solar activity and TSI on multiple time scales. Scafetta proposed
a physical mechanism that may explain how the small energy dissipated
by the gravitational tides may be significantly amplified up to a
4-million factor by activating a modulation of the solar nuclear fusion
rate \citep{Scafetta2012d}. However, the additional presence of theoretical
synodic cycles and an 11-year solar cycle modulation of the sub-annual
TSI variability also suggest electromagnetic planet-sun interactions
that could more directly drive the solar outer regions. Thus, if the
planets are modulating solar activity as our analysis suggests, the
solar response to planetary forcing would be complex and would non-linearly
depend on the 11-year solar cycle. Further research is required to
investigate the physical mechanisms of planetary-solar interactions
and construct models capable of simulating and predicting solar activity
and TSI variability.

\section*{Appendix}

The data were downloaded from here:

ACRIMSAT/ACRIM3: \href{http://acrim.com/RESULTS/data/acrim3/daya2sddeg_ts4_Nov_2013_hdr.txt}{http://acrim.com/RESULTS/data/acrim3/daya2sddeg\_{}ts4\_{}Nov\_{}2013\_{}hdr.txt}

SOHO/VIRGO: \href{ftp://ftp.pmodwrc.ch/pub/data/irradiance/virgo/TSI/virgo_tsi_d_v6_002_1302.dat}{ftp://ftp.pmodwrc.ch/pub/data/irradiance/virgo/TSI/virgo\_{}tsi\_{}d\_{}v6\_{}002\_{}1302.dat}

SORCE/TIM: \href{http://lasp.colorado.edu/data/sorce/tsi_data/daily/sorce_tsi_L3_c24h_latest.txt}{http://lasp.colorado.edu/data/sorce/tsi\_{}data/daily/sorce\_{}tsi\_{}L3\_{}c24h\_{}latest.txt}

SSN: \href{http://sidc.oma.be/silso/DATA/dayssn_import.dat}{http://sidc.oma.be/silso/DATA/dayssn\_{}import.dat}

\section*{Acknowledgment: }

The National Aeronautics and Space Administration supported Dr. Willson
under contracts NNG004HZ42C at Columbia University, Subcontracts 1345042
and 1405003 at the Jet Propulsion Laboratory.

\newpage{}

\begin{table}
\begin{tabular}{cccccc}
\hline
Cycle  & Type & P (day)  & P (year)  & min (year) & max (year)\tabularnewline
\hline
Me & $\nicefrac{1}{2}$ orbital & $44\pm0$ & $0.1205\pm0.000$ & $0.1205$ & $0.1205$\tabularnewline
\hline
Me – Ju & spring & $45\pm9$ & $0.123\pm0.024$ & $0.090$ & $0.156$\tabularnewline
\hline
Me – Ea  & spring & $58\pm10$ & $0.159\pm0.027$ & $0.117$ & $0.189$\tabularnewline
\hline
Me – Ve & spring & $72\pm8$ & $0.198\pm0.021$ & $0.156$ & $0.219$\tabularnewline
\hline
Me & orbital & $88\pm0$ & $0.241\pm0.000$ & $0.241$ & $0.241$\tabularnewline
\hline
Me – Ju  & synodic & $90\pm1$ & $0.246\pm0.002$ & $0.243$ & $0.250$\tabularnewline
\hline
Ea & $\nicefrac{1}{4}$ orbital & $91\pm3$ & $0.25\pm0.000$ & $0.250$ & $0.250$\tabularnewline
\hline
Ve & $\nicefrac{1}{2}$ orbital & $112.5\pm0$ & $0.3075\pm0.000$ & $0.3075$ & $0.3075$\tabularnewline
\hline
Me – Ea  & synodic & $116\pm9$ & $0.317\pm0.024$ & $0.290$ & $0.354$\tabularnewline
\hline
Ve – Ju & spring & $118\pm1$ & $0.324\pm0.003$ & $0.319$ & $0.328$\tabularnewline
\hline
Ea & $\nicefrac{1}{3}$ orbital & $121\pm7$ & $0.333\pm0.000$ & $0.333$ & $0.333$\tabularnewline
\hline
Me – Ve  & synodic & $145\pm12$ & $0.396\pm0.033$ & $0.342$ & $0.433$\tabularnewline
\hline
Ea & $\nicefrac{1}{2}$ orbital & $182\pm0$ & $0.500\pm0.000$ & $0.5$ & $0.5$\tabularnewline
\hline
Ea – Ju & spring & $199\pm3$ & $0.546\pm0.010$ & $0.531$ & $0.562$\tabularnewline
\hline
Ve & orbital & $225\pm0$ & $0.615\pm0.000$ & $0.241$ & $0.241$\tabularnewline
\hline
Ve – Ju  & synodic & $237\pm1$ & $0.649\pm0.004$ & $0.642$ & $0.654$\tabularnewline
\hline
Ve – Ea  & spring & $292\pm3$ & $0.799\pm0.008$ & $0.786$ & $0.810$\tabularnewline
\hline
Ea & orbital & $365.25\pm0$ & $1.000\pm0.000$ & $1.000$ & $1.000$\tabularnewline
\hline
Ea – Ju  & synodic & $399\pm3$ & $1.092\pm0.009$ & $1.082$ & $1.104$\tabularnewline
\hline
Ea – Ve & synodic & $584\pm6$ & $1.599\pm0.016$ & $1.572$ & $1.620$\tabularnewline
\hline
\end{tabular}\caption{List of the major theoretical expected harmonics associated with planetary
orbits within 1.6 year period. $P$ is the period. Mercury (Me), Venus
(Ve), Earth (Ea), Jupiter (Ju). If $P_{1}$ and $P_{2}$ are the periods,
the synodic period is $P_{12}=1/|1/P_{1}-1/P_{2}|$, and the spring
period is half of it. The variability is based on ephemeris calculations.
From \citet{ScafettaW2013b}.}
\end{table}
\begin{table}
\begin{tabular}{ccccc}
\hline
Cycle  & Type & P (year)  & Type & P (year) \tabularnewline
\hline
Me – Ne & spring & $0.1206$ & synodic & $0.2413$\tabularnewline
\hline
Me – Ur  & spring & $0.1208$ & synodic & $0.2416$\tabularnewline
\hline
Me – Sa & spring & $0.1215$ & synodic & $0.2429$\tabularnewline
\hline
Me – Ma & spring & $0.1382$ & synodic & $0.2763$\tabularnewline
\hline
Ve – Ne & spring & $0.3088$ & synodic & $0.6175$\tabularnewline
\hline
Ve – Ur & spring & $0.3099$ & synodic & $0.6197$\tabularnewline
\hline
Ve – Sa & spring & $0.3142$ & synodic & $0.6283$\tabularnewline
\hline
Ve – Ma & spring & $0.4571$ & synodic & $0.9142$\tabularnewline
\hline
Ea – Ne & spring & $0.5031$ & synodic & $1.006$\tabularnewline
\hline
Ea – Ur & spring & $0.5060$ & synodic & $1.0121$\tabularnewline
\hline
Ea – Sa & spring & $0.5176$ & synodic & $1.0352$\tabularnewline
\hline
Ea – Ma & spring & $1.0676$ & synodic & $2.1352$\tabularnewline
\hline
Ma & $\nicefrac{1}{2}$ orbital & $0.9405$ & orbital & $1.8809$\tabularnewline
\hline
Ma – Ne & spring & $0.9514$ & synodic & $1.9028$\tabularnewline
\hline
Ma – Ur & spring & $0.9621$ & synodic & $1.9241$\tabularnewline
\hline
Ma – Sa & spring & $1.0047$ & synodic & $2.0094$\tabularnewline
\hline
Ma – Ju & spring & $1.1178$ & synodic & $2.2355$\tabularnewline
\hline
Ju & $\nicefrac{1}{2}$ orbital & $5.9289$ & orbital & $11.858$\tabularnewline
\hline
Ju – Ne & spring & $6.3917$ & synodic & $12.783$\tabularnewline
\hline
Ju – Ur & spring & $6.9067$ & synodic & $13.813$\tabularnewline
\hline
Ju – Sa & spring & $9.9310$ & synodic & $19.862$\tabularnewline
\hline
Sa & $\nicefrac{1}{2}$ orbital & $14.712$ & orbital & $29.424$\tabularnewline
\hline
Sa – Ne & spring & $17.935$ & synodic & $35.870$\tabularnewline
\hline
Sa – Ur & spring & $22.680$ & synodic & $45.360$\tabularnewline
\hline
Ur & $\nicefrac{1}{2}$ orbital & $41.874$ & orbital & $83.748$\tabularnewline
\hline
Ur – Ne & spring & $85.723$ & synodic & $171.45$\tabularnewline
\hline
Ne & $\nicefrac{1}{2}$ orbital & $81.862$ & orbital & $163.72$\tabularnewline
\hline
Me – (Ju – Sa) & spring & $0.122$ & synodic & $0.244$\tabularnewline
\hline
Me – (Ea – Ju) & spring & $0.155$ & synodic & $0.309$\tabularnewline
\hline
Ve – (Ju – Sa) & spring & $0.317$ & synodic & $0.635$\tabularnewline
\hline
Ea – (Ju – Sa) & spring & $0.527$ & synodic & $1.053$\tabularnewline
\hline
Ve – (Ea – Ju) & spring & $0.704$ & synodic & $1.408$\tabularnewline
\hline
\end{tabular}\caption{List of additional average theoretical expected harmonics associated
with planetary orbits: Mercury (Me), Venus (Ve), Earth (Ea), Mars
(Ma), Jupiter (Ju), Saturn (Sa), Uranus (Ur), Neptune (Ne). If $P_{1}$
and $P_{2}$ are the periods, the synodic period is $P_{12}=1/|1/P_{1}-1/P_{2}|$,
and the spring period is half of it (\citet{ScafettaW2013b}). The
last five rows report additional spring and synodic periods of Mercury,
Venus and Earth relative to the synodic periods of Jupiter and Saturn,
and Earth and Jupiter. The latter periods are calculated using using
the three synodic period equation: $P_{1(23)}=1/|1/P_{1}-|1/P_{2}-1/P_{3}||$
.}
\end{table}
\begin{table}
\begin{tabular}{ccccc}
Cycle  & Type & P (day) & P (year)  & color\tabularnewline
\hline
Sun  & equ-rot & 24.7 & 0.0676  & black\tabularnewline
\hline
Sun – Ju & equ-rot & 24.8 & 0.0679 & red\tabularnewline
\hline
Sun – Ea & equ-rot & 26.5 & 0.0726 & red\tabularnewline
\hline
Sun – Ea & Car-rot & 27.3 & 0.0747 & blue\tabularnewline
\hline
Sun – Ve & equ-rot & 27.8 & 0.0761 & red\tabularnewline
\hline
Sun – Me & equ-rot & 34.3 & 0.0940 & red\tabularnewline
\hline
2/5 Me  & resonance & 35.2 & 0.0964 & green\tabularnewline
\hline
\end{tabular}\caption{Solar equatorial (equ-) and Carrington (Car-) rotation cycles relative
to the fixed stars and to the four major tidal planets calculated
using the synodic period equation: $P_{12}=1/|1/P_{2}-1/P_{2}|$,
where $P_{1}=24.7$ days is the sidereal equatorial solar rotation
and $P_{2}$ the orbital period of a planet. Last row reports the
2/5 Mercury's orbital resonance. The last column reports the color
of the arrows shown in Figure 6.}
\end{table}

\end{document}